\documentclass[twocolumn,showpacs,preprintnumbers,amsmath,amssymb,prl,reprint]{revtex4}

\usepackage{dcolumn}
\usepackage{bm}

\usepackage{amsmath}
\usepackage{amsfonts}   
\usepackage{graphics}  
\usepackage{psfrag} 
\usepackage{graphicx} 
\usepackage{epsfig}    
\usepackage{rotating}  
 \usepackage[latin1]{inputenc}
\usepackage[usenames,dvipsnames]{color}
 \usepackage{rotating}

\usepackage{color}
\usepackage{epsfig}
\definecolor{darkgreen}{rgb}{0,0.5,0} 
\definecolor{violet}{rgb}{0.5,0,0.5}
\definecolor{orange}{rgb}{0.2,0.5,0.5}

\begin{document}

\preprint{}

\title{Evolutionary Fitness in Variable Environments
}

\author{Anna Melbinger and Massimo Vergassola}
\affiliation{University of California San Diego, Department~of Physics, 9500~Gilman Drive, La Jolla, CA 92093}


\date{\today}
\begin{abstract}
One essential ingredient of evolutionary theory is the concept of fitness as a measure for a species' success in its living conditions. Here, we quantify the effect of 
 environmental fluctuations onto fitness by analytical calculations on a general evolutionary model and by studying  corresponding individual-based microscopic models. We demonstrate that not only  larger growth rates and viabilities, but also reduced sensitivity to environmental variability substantially increases the fitness. Even for neutral evolution, variability in the growth rates plays the crucial role of strongly reducing the expected fixation times. Thereby, environmental fluctuations constitute a mechanism to account for the effective population sizes inferred from genetic data that often are much smaller than the census population size.
\end{abstract}

\pacs{87.23.-n,02.50.Ey}
\maketitle

Spencer's famous expression ``survival of the fittest''~\cite{Spencer} provides an appealing  short summary of Darwin's concept of evolution~\cite{Darwin,Wallace:1858}.  However, it leaves aside 
a very difficult  yet important aspect namely identifying the factors determining the fitness of a species~\cite{Metz:1992,Ariew:2004}\,: fittest individuals are by definition prevailing but the reasons facilitating their survival are not obvious. Besides the difficulties arising due to the genotype phenotype mapping causing complex fitness landscapes~\cite{Visser:2014}, also ecological factors like population structure and composition additionally complicate the issue. Therefore traditional fitness concepts solely based on growth rates and viability were extended by frequency-dependent~\cite{MaynardSmith:1973} or inclusive fitness approaches~\cite{Hamilton:1964}. Another important factor for the success  of a certain trait, is a non-constant environment influencing birth/death rates~\cite{Gillespie:1973b,Takahata:1975,Frank:1990,May:1973,Haccou:1994,Yoshimura:1996,Kussell:2005,Orr:2007,Chevin:2010, Riviore:2014}. How variable environmental conditions affect evolutionary strategies like phenotypic heterogeneity or bet-hedging has been extensively studied, see \emph{e.g.}~\cite{Schaffer:1974,Kussell:2005b, Kussell:2005,Acer:2008, Beaumont:2009,Pintu:2014}, yet the consequences of  fluctuating reproduction rates and their interplay with demographic fluctuations were not fully elucidated. 

Here, we quantitatively investigate the impact of variable environments on the fitness. In contrast to other models dealing with variable environmental conditions, we do not study which strategy is optimally suited to cope with such changing environments but focus on the consequences of fluctuating reproduction rates. In particular, we show that an individual's sensitivity to environmental changes contributes substantially to its fitness: A reduced sensitivity increases the fitness and may compensate for large disadvantages in the average reproduction rate. We also find that fluctuating environments influence neutral evolution where they can cause much quicker fixation times than expected. These effects are relevant as constant environmental conditions are the exception rather than the norm; for instance, the availability of different nutrients, the presence of detrimental substances and other external factors like temperature, all strongly  influence reproduction/survival and occur on a broad range of time scales  \cite{Mustonen:2007}. 

To understand the impact of fluctuating environmental conditions, we first consider rapidly changing environments in an evolutionary process based on birth and death events similar to~\cite{May:1973}. The dynamics is described by the following stochastic differential equations\,:
\begin{align}
\dot N_S\!=\!\left(\!\nu_S\!-\!\gamma\!\frac{N}{K}\!\right)\!N_S\!+\!N_S\sigma_S\xi_S\!+\!\sqrt{\!N_S(\nu_S\!+\!\gamma\!\frac{N}{K}})\mu_S.
\label{eq:Langevin}
\end{align}
The influence of environmental variability is modeled as white noise acting on the growth rate, $\nu_S$, of a trait of type $S$: $\nu_S+\sigma_S\xi_S$, where $\langle \xi_S(t)\xi_{S}(t')\rangle=\delta(t-t')$ and $\sigma_S$ is the Standard Deviation (STD) of the noise. Death rates are assumed to be constant and identical for all traits~\cite{FN}.  Population growth is bounded and therefore  death rates increase with the total population size, $N\!=\!\sum_S N_S$ where $N_S$ is the number of $S$-type individuals~\cite{Verhulst}. This  may account for density-dependent ecological factors such as limited resources or metabolic waste products accumulating at high population sizes. For specificity, we choose $\gamma N/K$ as functional form where $K$ is the carrying capacity scaling the maximal number of individuals and $\gamma$ sets the rate  of death events. 
Beside environmental noise, demographic fluctuations arising from the stochastic nature of the birth-death dynamics yield the term $\sqrt{N_S(\nu_S\!+\gamma N/K)}\mu_S$, where $\mu_S$ is $\delta$-correlated noise, $\langle \mu_S(t)\mu_{S}(t')\rangle\!=\!\delta(t-t')$, with a variance given by the sum of reaction rates~\cite{Gillespie}. Both multiplicative noise terms in Eq.~\eqref{eq:Langevin} are interpreted in the Ito sense \cite{VanKampen:2001}. Note that environmental noise is linearly multiplicative in $N_S$, which is crucial for our results.  

Let us now consider the Fokker-Planck equation (FPE) associated to Eq.~\eqref{eq:Langevin}, which we will use to derive fixation probabilities and times. We shall carry out further analysis for two different traits $S\in\{\text{\bf1,2}\}$; generalizations are straightforward. The transformation of Eq.~\eqref{eq:Langevin} to a FPE, depends on the correlation level of the environmental noise acting on distinct traits~\cite{Gillespie:1996}. While demographic noise for different traits is always uncorrelated, the same environmental noise can affect multiple traits, e.g. if both traits feed from the same nutrients whose abundance fluctuates. We keep the following analysis quite general by introducing the correlation coefficient $\epsilon$\,: for  $\epsilon\!=\!0$ environmental noise is  uncorrelated, $\langle \xi_1\xi_2\rangle=0$, while $\epsilon\!=\!\pm 1$ for $\xi_1\!=\!\pm \xi_2$. The resulting FPE is
 \begin{align}
 &\frac{\partial P(\!N_1,\!N_2,t)}{\partial t}\!=\!\epsilon \partial^2_{1,2}\sigma_1\sigma_2N_1N_2P - \sum_{i}\partial_{i}\left[\left(\nu_i\!-\!\gamma\frac{N}{K}\right)\!N_i P\right]\nonumber \\
&+\frac{1}{2}\sum_i
 \partial^2_{i}\left\{ \left[(N_i\sigma_i)^2+ N_i\left(\nu_i+\gamma\frac{N}{K}\right)\right]P\right\}\,,
 \end{align}
 where $\partial_i\equiv \partial_{N_i}$. To uncover the influence of environmental noise on the evolutionary dynamics, the relative abundances seem the natural observables. Therefore, we change variables to the fraction $x=\frac{N_1}{N_1+N_2}$ and the total number of individuals $N=N_1+N_2$. The FPE for $x$ and $N$ can be simplified exploiting  the fact that selection, $s=\nu_1-\nu_2$, is much slower than population growth $\nu_1x+\nu_2(1-x)$. Therefore, we integrate over the total population size $N$, considering the FPE for the marginal distribution $P(x)=\int_0^\infty P(x,N)\,dN$, and employing $N\gg1$, see SM~\cite{SM}. The resulting one-dimensional FPE reads\,:
 \begin{align}
 &\frac{\partial P(x,t)}{\partial t} = \partial_x\left\{ \left[-s\!-\!\sigma_2^2(1\!-\!x)\!+\!\sigma_1^2x\!+\!\epsilon\sigma_1\sigma_2\left(1-2x\right)\right]Q\right\} \nonumber \\
&+\!\partial^2_x\!\left\{\!\left[\frac{\sigma_1^2\!-\!2\epsilon\sigma_1\sigma_2\!+\!\sigma_2^2}{2}x(1\!-\!x)\!+\!\!\frac{\gamma}{K}\right]\! Q\right\}\!\!\equiv\! {\cal L}P(x,t), \label{eq:FP}
\end{align}
where $Q\equiv x(1-x)P(x,t)$ and the last equality defines the Fokker-Planck operator ${\cal L}$ needed in the sequel. In Ref.~\cite{Takahata:1975} a similar FPE was derived for the special case $\sigma_1=\sigma_2$. For $\sigma_1\!=\!\sigma_2=0$, the drift term reduces to the well-know expression $-s\partial_x Q$ favoring the trait with a higher growth rate~\cite{Kimura}. Note that the variability in the growth rates affects both the diffusion term and the drift, which is due to the multiplicative nature of the environmental noise. 

For simplicity, we discuss the case of {\it different environmental sensitivity}, defined by $\sigma_1=\Delta$ and $\sigma_2=0$, \emph{i.e.} only the reproduction rate of the first trait depends on the environment. The drift is then proportional to $\alpha(x)=(s-\Delta^2 x)x(1-x)$ independent of $\epsilon$. If $s<0$, \emph{i.e.} the second trait  with a smaller variability in its birth rate is also faster in reproducing, the evolutionary dynamics does not change qualitatively compared to $\Delta=0$. Conversely, if $s>0$, the situation changes dramatically\,:   the growth rate favors trait {\bf1} while the variability term favors trait {\bf2}. This leads to a stable fixed point $x^*=\frac{s}{\Delta^2}$ for $s<\Delta^2$ (for $s>\Delta^2$ variability is not sufficient to prevent extinction of trait {\bf2}). Such a dynamics can be interpreted as  frequency-dependent fitness function. However, the frequency-dependence arises here from environmental noise and not from 
a pay-off matrix~\cite{Maynard} as in standard evolutionary game theory. 
\begin{figure}[t]
\includegraphics[width=\columnwidth]{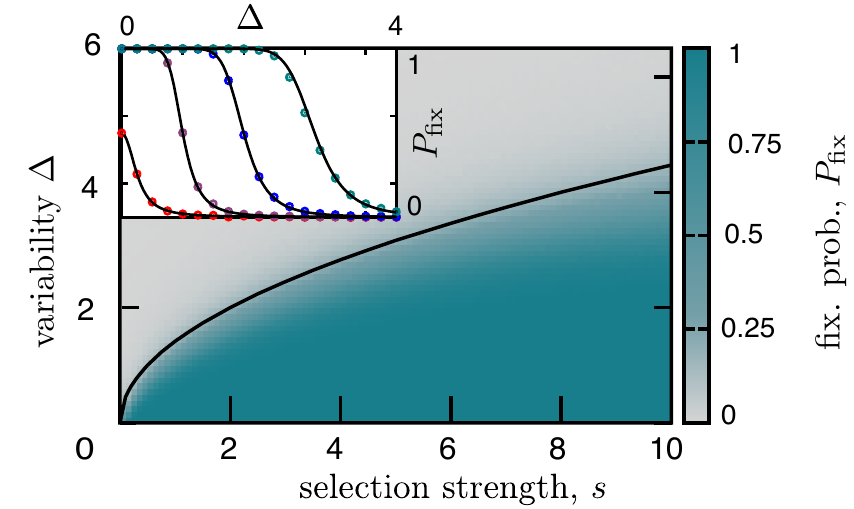}
\caption{Fixation probability, $P_\text{fix}$, depending on selection strength, $s$, and variability $\sigma_1\!=\!\Delta$ according to Eq.~\eqref{eq:Langevin}. Other parameters are $\nu_1\!=\!10$, $\sigma_2=0$, $\gamma=1$ and $K=100$. The black line indicates the parabola $s=\Delta^2/2$, which is our prediction for $P_\text{fix}=0.5$. The inset shows cuts for exemplary values of $s=\{0,0.5,2,5\}$ in $\{$red, violet, blue, green$\}$. \label{fig:A}}
\vspace{-0.6cm}
\end{figure}

Even though environmental variability causes a drift term favoring the traits which is less sensitive to environmental changes~\cite{Frank:2011}, the interplay between drift and diffusion term has to be understood to predict the evolutionary outcome. This is even more important as for the particular situation discussed here, the environmental contribution to the drift caused by  $\sigma_S$ is intrinsically connected to the diffusion term. Therefore we study the fixation probability, \emph{i.e.} the probability that trait {\bf 1} fixates or trait {\bf 2} goes extinct. This quantity can be calculated by solving the backward FPE, $0=\mathcal{L}^\dag_{x_0}P_\text{fix}(x_0)$ for the boundary conditions $P_\text{fix}(0)=0$ and $P_\text{fix}(1)=1$. The solutions is given by (for details see SM),
\begin{align}
P_\text{fix}(x)=\!\frac{\!1\!-\!\exp\!\left\{{\zeta\! \left[\text{Tanh}^{-1}\alpha \!+\!\text{Tanh}^{-1}\alpha (2 x-1)\right]}\right\}}{1-\exp{\left\{2\zeta\, \text{Tanh}^{-1}\alpha\right\}}},
\label{eq:fix_prob}
\end{align}
with $\beta=\sqrt{K(\sigma_1^2-2\epsilon\sigma_1\sigma_2+\sigma_2^2)/\gamma}$, $\alpha\equiv \beta/\sqrt{8+\beta^2}$ and $\zeta\equiv 2K \left(\sigma_1^2-\sigma_2^2-2 s\right)/(\beta\gamma \sqrt{8+\beta^2})\,.$
In Fig.~\ref{fig:A} we show the fixation probability  for different values of $s$ and $\sigma_1=\Delta$ ($\sigma_2=0$). Results are obtained by the numerical solution of Eq.~\eqref{eq:Langevin}, \emph{i.e}. before marginalization on $N$. The parabola  $s=\Delta^2/2$ (Fig.~\ref{fig:A} black line) defined by $x^*= 0.5$ (or $P_\text{fix}=0.5$), separates the regions where 
one of the two traits is predominant\,: in the grey (green) area, the smaller variability (growth rate) dominates, respectively.   The general case of both species having variable birth rates yields analogous results: a selection advantage for the trait with less variability. 
In the inset, the fixation probability depending on $\Delta$ is compared to the analytic solution (Eq.~\eqref{eq:fix_prob}) for the four values $s=\{0,0.5,2,5\}$. Both plots demonstrate the advantage of the less variable trait. For strong environmental variations it is then beneficial for a species to minimize its sensitivity to those variations rather than optimizing its growth rate. Interestingly one can interpret this result in the context of game theory: Decreasing the sensitivity to environmental changes also means to optimize the worst case scenario outcome: The average birth rate is the least reduced when the variability is small. In game theory, this corresponds to the MaxiMin strategy which was  shown to be very successful in many fields as finance, economy or behavioral psychology~\cite{Neumann:1944,Owen:1995}. In the context of evolutionary dynamics another example of a MaxiMin strategy was proposed for bacterial chemotaxis where bacteria move 
move so as to optimize their minimal uptake of chemoattractants~\cite{Celani2009}.

Besides contributing to the fitness, environmental variability also influences fixation probability and time in the case of  \emph{neutral evolution}, \emph{i.e.} $\nu_1\!=\!\nu_2$ and $\sigma_1\!=\!\sigma_2=\sigma$. Such analysis is of great interest, as evolution is often studied by investigating how neutral mutations evolve over time. In recent years fast-sequencing techniques made huge amounts of data available, see, \emph{e.g.}, \cite{Shendure:2008}, which is now analyzed and interpreted by comparison to evolutionary models as the Moran or Fisher-Wright models~\cite{Blythe:2007}.
While the correlation parameter does not qualitatively influence results discussed so far, it plays an important role for neutral evolution. For fully correlated noise, $\epsilon=1$, Eq.~\eqref{eq:FP} is the same as for $\sigma_1=\sigma_2=0$ and thereby correspond to the ones obtained for no environmental noise, extinctions are solely driven by demographic fluctuations and well-known results apply~\cite{Kimura}. In contrast, for all other values of $\epsilon$, including uncoupled noise $\epsilon=0$, the dynamics differs in two major respects. First, the drift term $-\sigma^2\left(1-\epsilon\right)\partial_x(1-2x)x(1-x)P(x)$ does not vanish and corresponds to a stable fixed point at $x^*=0.5$. Second, the diffusion term consists of demographic $\frac{\gamma}{K}x(1-x)$ and environmental fluctuations $\left(1-\epsilon\right)\sigma^2x^2(1-x)^2$.  As the drift suppresses extinction events while a larger diffusion term favors them, a more detailed analysis is required to grasp the evolutionary outcome. 

\begin{figure}[t]
\centering
\includegraphics[width=\columnwidth]{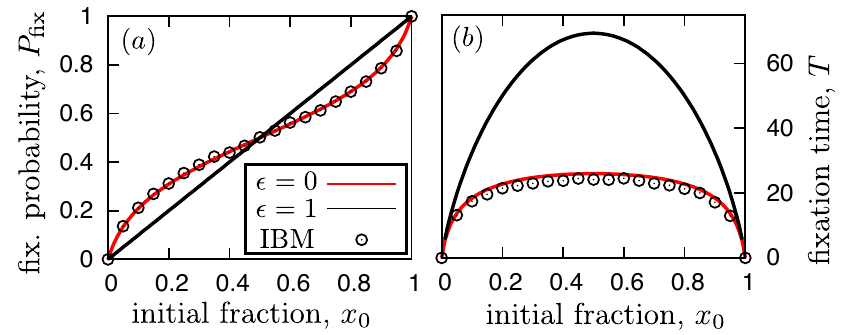}
\caption{Fixation probability (panel a) and  time (panel b) in the neutral case. Solid lines indicate 
analytical results for the two typical cases 
of perfectly correlated and uncorrelated noises $\epsilon=1$ and $\epsilon=0$.
Parameters are\,: $\nu_1=\nu_2=10$, $K=100$, $\sigma_1\!=\!\sigma_2\!=\!0.5$ and $\gamma=1$. Dots are simulations of the IBM. Additional parameters are $m=1,~\phi_1\!=\!\phi_2\!=\!10$, $  \omega_1\!=\!\omega_2=5$, $\tau\!=\!0.01$, $\langle E\rangle\!=\!0$, $\alpha_1=\alpha_2=1$, and $\text{Var}[E]\!=\!100$. \label{fig:2}}
\end{figure}

\begin{figure}[b]
\centering
\includegraphics[width=\columnwidth]{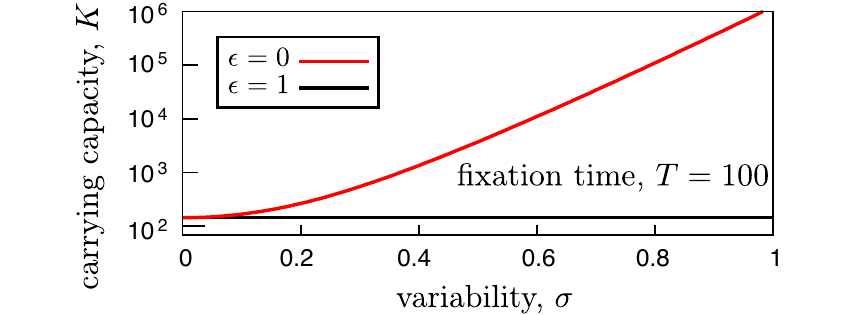}
\caption{Reduction of the effective population size due to environmental noise in the neutral case for $x_0=0.5$. Using \eqref{eq:fix_time}, we plot the values of the population size $K$ and noise 
$\sigma$ that lead to the fixation time $T=100$ for $\gamma=1$. The black line corresponds to $\tilde\sigma=0$ (see \eqref{eq:fix_time}), i.e. either perfectly correlated noise or no environmental noise. In the presence of environmental  noise, the values of $K$  are systematically higher and  increase  several orders of magnitude even for moderate noise levels.
 \label{fig:3}}
\end{figure}

Due to the stable fixed point, the fixation probability  qualitatively differs from the linear dependence $P_\text{fix}^{\epsilon=1}=x_0$
 which holds for constant or uncorrelated environments. In Fig~\ref{fig:2}a), typical solutions for  $\epsilon=1$ and $\epsilon=0$ are shown which clearly demonstrated the ensuing s-shape for the uncorrelated case ($\epsilon=0$). 
Another important quantity, the extinction time, $T(x_0)$, also obeys a backward FPE, $-1=\mathcal{L}^{\dag}_{x_0}T(x_0)$. Employing the boundary conditions $T(0)=T(1)=0$, the fixation time in the neutral case ($s=0$ and $\sigma_1=\sigma_2=\sigma$) can be calculated~\cite{SM}\,:
\begin{align}
T&=\frac{1}{C\tilde\sigma^2}\left[
\ln\frac{1\!-\!\Gamma_{\!+}(1\!-\!x_0)}{1\!-\!\Gamma_{\!-}(1\!-\!x_0)}\ln(1\!-\!x_0)\!+\!\ln\!\frac{1\!-\!\Gamma_{\!+}x_0}{1\!-\!\Gamma_{\!-}x_0}\ln x_0\right.\nonumber\\ &\left.+F_{\Gamma_+}(x_0)-F_{\Gamma_-}(x_0)\right]\,,
\label{eq:fix_time}
\end{align}
where $\tilde\sigma=(1-\epsilon)\sigma$, $C=\sqrt{1+4\gamma/(K\tilde\sigma^2)}$, $\Gamma_\pm\!=\!2/(1\pm C)$,  the function $F_{\Gamma}(x)\equiv \text{Li}_2(\Gamma(1\!-\!x))+\text{Li}_2(\Gamma x)-\text{Li}_2(\Gamma)$ and $\text{Li}_n$ is the polylogarithm. The result for $\epsilon<1$  differs again from the non-fluctuating/fully correlated scenario, $T^{\epsilon\!=\!1}=-K/\gamma[x_0\ln(x_0)+(1-x_0)\ln(1-x_0)]$ (see Fig.~\ref{fig:2}b). Fluctuating environments decrease the fixation times for all initial conditions. This has a crucial consequence: when measuring extinction times and comparing them to standard models without environmental fluctuations, one can only explain large diffusion constants by small population sizes. Indeed, it is often found that effective population sizes  are much smaller than the census population sizes~ \cite{Charlesworth:2009ph}. Fig.~\ref{fig:3} shows that conspicuous orders-of-magnitude reductions in the population size set in already at moderate levels of environmental noise. Amongst other explanations this could account for a difference between effective and census population size. In other words as long as the level of environmental noise and the correlation level of its influence on different growth rates is not known, the effective population size can only be interpreted as a lower bound for the census population size.

To further investigate the impact of variable environmental conditions, we introduce an exemplary Individual Based Model (IBM). In particular, the IBM serves as a proof of principle that linear multiplicative noise can be realistically expected and enables us to study the effect of such noise beyond the white noise regime. Importantly, our results presented above hold for any microscopic model whose macroscopic representation is given by Eq.~\eqref{eq:Langevin}, \emph{i.e.} where noise in the birth or death rates is linearly multiplicative. In specific scenario discussed here, the reproduction rate of an individual, $i$, at time $t$, depends a priori on the history of environmental conditions  experienced during its lifetime $t^i_\text{life}=[t_0^i,t]$, where $t_0^i$ is the time of birth. This could for example account for the level of nutrients or detrimental substances that individuals are exposed to.
Following~\cite{Melbinger:2010, Cremer:2011a}, our model is based on independent birth and death rates, now depending on the environmental variations subsumed in the scalar value $E$. The number of environments experienced by an individual, $i$, is denoted as $M^i$ and their values are contained in a vector $\vec E^{i}=(E_1^{i},E_2^{i},...,E^i_{M^{i}})$. For the fully correlated case ($\epsilon\!=\!1$), $E$ is the same for both traits while two different values are drawn  for $\epsilon\!=\!0$. Environmental conditions change stochastically at rate $1/\tau$ and are distributed according to a distribution, $p(E)$, with mean  $\langle E\rangle $ and variance $\text{Var}[E]$.  

We first consider a constant environment $E$. The average \emph{instantaneous growth rate} $\lambda_S(E)$ is assumed to be a positive, monotonically increasing function of $E$~\cite{FN2}. In particular, we consider the sigmoidal function\,:
  \begin{align}
 \lambda_S(E)=\phi_S+\omega_S\tanh\left(\alpha_S E/2\right)\,,
 \label{eq:lambda}
 \end{align}
 with $\phi_S$ the ordinate of the inflection point, $\omega_S\le \phi_S$ the maximal deviation from it, and $\alpha_S$ scales the sensitivity to environmental changes. 

Let us now consider changing environments and individuals whose current growth rate memorizes previously experienced environments. The reproduction rate $\Gamma^i_\text{repr,S}$ of an individual, $i$, of type $S$ now depends on the whole vector, $\vec E^{i}$.
For concreteness, we  assume that the rate is 
\begin{align}
\Gamma^i_\text{repr,S}=\frac{1-m~~~}{1-m^{M^{i}}}\sum_{j=1}^{M^{i}}m^{j-1}\lambda(E^i_j),
\label{eq:Gamma}
\end{align} 
where the memory parameter $m\in[0;1]$ defines the influence of previously experienced environments upon an individual's growth rate. For $m=0$ only the current environment sets $\Gamma^i_\text{repr,S}=\lambda(E^i_{M^i})$, while for $m\rightarrow1$ all experienced environments, $M^i$, have the same influence in the arithmetic mean $\Gamma^i_\text{repr,S}=\frac{1}{M^i}\sum_{k\leq M^i} \lambda^k_S( E_k^{i})$. Independent of $m$ we assume that offsprings lose memory at the time of reproduction. Bounded growth is modeled by death rates $\Gamma^i_\text{death,S}=\gamma N/K$.

 To compare the results of the microscopic individual based model to the effective stochastic model, Eq.~\eqref{eq:Langevin}, the parameters of both models have to be mapped.  For simplicity let us consider the case $\langle E\rangle=0$ and a symmetric distribution $p(E)$ throughout the following discussion. Since death rates are constant, there is a direct correspondence between them in the Langevin and the IBM. For birth rates and their STDs the situation is more intricate as we discuss hereafter. For the no-memory case ($m=0$) an exact mapping is obtained~\cite{SM}: For strong fluctuations, $\alpha_S^2\text{Var}[E]\gg 1$, the mean of the growth rate $\nu_S$ and STD of the noise $\sigma_S$ in Eq.~\eqref{eq:Langevin} are given by\,:
\begin{align}
\nu_S(m=0)=\phi_S+\omega^2\tau,~~
\sigma_S(m=0)=\omega\sqrt{2\tau}.
 \label{eq:mapping0}
\end{align}
Note that the variability in the growth rate not only results in $\sigma_S>0$, but also influences the average reproduction rate $\nu_S$. While for $m=0$ such a variability increases $\nu_S$, the second term of $\nu_S$ is reduced while $m$ increases till it changes sign (see 
SM~\cite{SM} for details). For instance, for $m=1$ the growth rate is approximately $\phi_S-\omega^2\tau$. Hence, the more variable trait has a disadvantage in the average reproduction rate in addition to the effects discussed above. For $m=1$, the approximation $\sigma_S(m=1)\approx\omega_S\sqrt{\tau}$ holds~\cite{SM}. Dependencies in this expression are intuited as follows. The number of environmental changes an individual experiences until the memory resets is of the order $M\sim t_\text{life}/\tau$, where $t_\text{life}\propto 1/\nu_S$ is the typical time for an individual to reproduce or die. As environmental changes are independent random events, the variance of the reproduction rates \eqref{eq:Gamma} is $\propto\omega_S^2/M$. The expression for $\sigma_S(m=1)$ is finally obtained noting that correlations in the noise extend over times $\sim t_\text{life}$ therefore it follows that the average reproduction rate $\nu_S$  drops out. 

  \begin{figure}[t]
 \centering
 \includegraphics[width=\columnwidth]{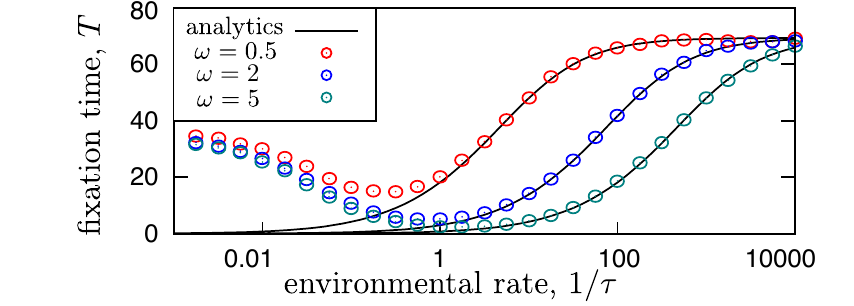}
 \caption{\label{fig:change_env}Comparison of the IBM and the Langevin model. We show the fixation time for neutral evolution for $x_0=0.5$ {\it vs} the environmental switching rate $1/\tau$. Dots correspond to the IBM [$m=0$] for different values of $\omega_1\!=\!\omega_2=\{0.5,2,5\}$ in red, blue and green. Black lines are analytic solutions [Eq.~\eqref{eq:fix_time} with  Eq.~\eqref{eq:mapping0}]. For quickly fluctuating environments both results are in good agreement whilst for large $\tau$ the white noise approximation fails. Other parameters are as in Fig.~\ref{fig:2}.
 }
 \end{figure}
 For a detailed comparison of the  IBM with the analytics derived in the first part of this paper, we simulate the IBM with a modified Gillespie algorithm updating reproduction rates after every environmental change~\cite{Gillespie}. As shown in Figs.~\ref{fig:2}a) and b), results for fixation probability and time, are in very good agreement with analytic solutions [Eqs.~\eqref{eq:fix_prob} and \eqref{eq:fix_time}]. In particular, the sigmoidal shape of the fixation probability is well reproduced by the IBM, supporting the existence and importance of linear multiplicative noise.

  Finally, the IBM enables us to study the environmental switching rate. This is of main interest as results obtained previously strictly only hold for very rapidly fluctuating environments. In Fig.~\ref{fig:change_env}, the dependency on $\tau$ of the extinction time in the neutral case for $x_0=0.5$ is shown for different $\omega_S$; see SM for results with $s\neq0$~\cite{SM}. The black lines correspond to Eq.~\eqref{eq:fix_time} mapped according to Eqs.~\eqref{eq:mapping0} and dots are obtained by stochastic simulations of the IBM for $m=0$. For $\tau<1$ both models are in very good agreement. This demonstrates that the white noise approximation is valid in a broad parameter range, where fluctuating environments substantially influence the evolutionary dynamics. 
  
In summary, we demonstrated  that environmental variability has crucial impact on evolutionary fitness. 
First, we quantified the role of reduced sensitivity to environmental changes and determined how it substantially increases the fitness. Second,  we showed that the timescale of extinction in the neutral case is strongly affected by environmental noise. That provides a mechanism to explain experimental observations of population sizes that are often much smaller than expected.  Finally, we investigated individual based models that generate the linear multiplicative noise considered here. It will be of interest to investigate how different forms of memory or time-dependent reproduction rates influence evolution and to integrate them with evolutionary game models.

\acknowledgements{We thank Jonas Cremer  for valuable  discussions and comments on the manuscript. AM acknowledges the German Academic Exchange Service (DAAD) for financial support.}

\renewcommand{\theequation}{S\arabic{equation}}
\renewcommand{\thefigure}{S\arabic{figure}}
\setcounter{figure}{0} 
\setcounter{equation}{0} 
\newpage
{\LARGE \bf  Supporting Information}\newline~\newline
In this Supporting Material, we provide more details on the calculations  leading to the one-dimensional Fokker-Planck equation (FPE), Eq.~(3)  main text. Moreover, we present calculations for the average fixation time and the extinction probability. We discuss the mapping of the individual based model (IBM) to the Langevin model. For the no-memory limit we present an analytic derivation of the mapping. For other parameter values we give heuristic arguments that are supported by additional data. Finally, we present results for non-neutral evolution investigating the regime in which the white noise approximation is an adequate description for the evolutionary process.
\section{Derivation of the one-dimensional Fokker-Planck equation}
In this Section, we provide details how the Langevin equations,
\begin{eqnarray}
\label{eq:Lang}
\dot N_1&=N_1(\nu_1-\gamma\frac{N}{K})+\sqrt{N_1\left(\nu_1+\gamma\frac{N}{K}\right)}\mu_1\!+\!\sigma_1N_1\xi_1\nonumber\\
\dot N_2&=N_2(\nu_2-\gamma\frac{N}{K})+\sqrt{N_2(\nu_2+\gamma\frac{N}{K})}\mu_2\!+\!\sigma_2N_2\xi_2
\end{eqnarray}
can be transformed into the one-dimensional FPE presented in the main text. From general results on stochastic processes (see \cite{Gnedenko}), it follows that the previous Langevin equation is associated to the following two-dimensional FPE\,: 
\begin{align}
 &\frac{\partial P(\!N_1,\!N_2,t)}{\partial t}\!=\!\epsilon \partial^2_{1,2}\sigma_1\sigma_2N_1N_2P - \sum_{i}\partial_{i}\left[\left(\nu_i\!-\!\gamma\frac{N}{K}\right)\!N_i P\right]\nonumber \\
&+\frac{1}{2}\sum_i
 \partial^2_{i}\left\{ \left[(N_i\sigma_i)^2+ N_i\left(\nu_i+\gamma\frac{N}{K}\right)\right]P\right\}\,,
 \label{FP2D}
 \end{align}
 where $\partial_i\equiv \partial_{N_i}$. The drift part is directly stemming from the non-fluctuating parts of the Langevin equations $N_S(\nu_S-\gamma N/K)$. 
Diffusion depends on the correlation level of the noises experienced by the two species. In particular, we have introduced the correlation coefficient $\epsilon\equiv \langle \xi_1\xi_2\rangle/\sqrt{\langle\xi_1^2\rangle\langle\xi_2^2\rangle}$. The case when the two noises are the same is given by $\epsilon=1$, when they are independent is  $\epsilon=0$ and when they are anti-correlated is $\epsilon=-1$.   
 
To study the evolutionary dynamics associated to Eq.~\eqref{FP2D}, the relative abundances are the natural choice of variables. Therefore, we transform the absolute abundances $N_1$ and $N_2$ to $x=\frac{N_1}{N_1+N_2}$ and $N=N_1+N_2$.
 To perform the change of variables, not only $N_1=xN$ and $N_2=(1-x)N$ have to be replaced, also the differential operators and the probability distribution have to be transformed. Ensuring that the latter is still normalized after change of variables, the Jacobian has to be introduced, $P(N_1,N_2)\rightarrow\frac{1}{N}P(x,N)$. The derivatives are given by, $\partial_{N_1}\rightarrow\frac{1-x}{N}\partial_x+\partial_N$ and $\partial_{N_2}\rightarrow-\frac{x}{N}\partial_x+\partial_N$.

After the change of variables, the FPE for $x$ and $N$ can now be further simplified exploiting  the fact that the time scale of selection, $s=\nu_1-\nu_2$, is much slower than the one of the population growth $\nu_1x+\nu_2(1-x)$~\cite{Cremer:2011a}. Therefore, we marginalize the FPE with respect to  the total population size $N$. Thereby,  the integrals $\int_0^{\infty}dN$ of $N$-derivative terms such as $\partial_N\bullet$ or $N\partial^2_N\bullet=\partial_N\left(N\partial_N\bullet\right)
-\partial_N\bullet$ vanish and the FPE simplifies to

 \begin{align}
 &\frac{\partial P(x,t)}{\partial t} = \partial_x\left\{ \left[-s\!-\!\sigma_2^2(1\!-\!x)\!+\!\sigma_1^2x\!+\!\epsilon\sigma_1\sigma_2\left(1-2x\right)\right]Q\right\} \nonumber \\
 &+\partial_x\left(\frac{s}{N} Q\right)\nonumber\\
&+\!\partial^2_x\!\left\{\!\left[\frac{\sigma_1^2\!-\!2\epsilon\sigma_1\sigma_2\!+\!\sigma_2^2}{2}x(1\!-\!x)\!+\!\!\frac{\gamma}{2K}+\frac{\nu_1-sx}{2N}\right]\! Q\right\}, \label{eq:FP1Db}
\end{align}
where $Q\equiv x(1-x)P(x,t)$. The drift term in the second line stemming from demographic fluctuations can be neglected as $N\gg1$ holds. To finally arrive at the one-dimensional FPE employed in the main text, we compute the steady state population size $N^*$. As the deterministic differential equation for $N$ is given by
\begin{align}
\dot N=N\left[x\nu_1+(1-x)\nu_2-\gamma\frac{N}{K}\right],\nonumber
\end{align}
the fixed point for the populations size is $N^*=K/\gamma [\nu_1x+\nu_2(1-x)]$. Employing that relation and the aforementioned condition $s\ll \nu_1x+\nu_2(1-x)$,  the last term in Eq.~\eqref{eq:FP1Db} can be simplified as $\frac{\nu_1-sx}{2N}\approx \frac{\gamma}{2K}$, which finally leads to the one-dimensional FPE in the main text:
 \begin{align}
 &\frac{\partial P(x,t)}{\partial t} = \partial_x\left\{ \left[-s\!-\!\sigma_2^2(1\!-\!x)\!+\!\sigma_1^2x\!+\!\epsilon\sigma_1\sigma_2\left(1-2x\right)\right]Q\right\} \nonumber \\
&+\!\partial^2_x\!\left\{\!\left[\frac{\sigma_1^2\!-\!2\epsilon\sigma_1\sigma_2\!+\!\sigma_2^2}{2}x(1\!-\!x)\!+\!\!\frac{\gamma}{K}\right]\! Q\right\}, \label{eq:FP1D}
\end{align}

\section{Fixation probability}

 In the following, we derive a general expression for the fixation probability. The calculations are analogous to the  procedure for the neutral case described in the body of the paper. To determine the fixation probability the following backward equation has to be solved,
\begin{align}
 0&=\!x\left(1-x\right) \bigg \{\!\left[s\!+\!\sigma_2^2(1\!-\!x)\!-\!\sigma_1^2x\!-\!\epsilon\sigma_1\sigma_2(1\!-\!2x)\right]\partial_x\nonumber \\ &+\!\left[\frac{\sigma_1^2\!-\!2\epsilon\sigma_1\sigma_2\!+\!\sigma_2^2}{2}x(1\!-\!x)+\!\frac{\gamma}{K}\right]\! \partial_x^2\bigg\}P_\text{fix}(x). \label{eq:FP}
\end{align}
Boundary conditions are $P_\text{fix}(0)=0$ and $P_\text{fix}(1)=1$. The solution to Eq.~\eqref{eq:FP} for the fixation probability is
\begin{align}
P_\text{fix}(x)=\frac{\!1\!-\!\exp\left\{{\zeta \left[\text{Tanh}^{-1}\alpha \!+\!\text{Tanh}^{-1}\alpha (2 x-1)\right]}\right\}}{1-\exp{\left\{2\zeta\, \text{Tanh}^{-1}\alpha\right\}}}\,,
\label{eq:fix_prob}
\end{align}
with 
\begin{eqnarray}
\beta&=&\sqrt{K(\sigma_1^2-2\epsilon\sigma_1\sigma_2+\sigma_2^2)/\gamma};\quad\alpha\equiv \frac{\beta}{\sqrt{8+\beta^2}};\nonumber \\\zeta&\equiv& \frac{2K \left(\sigma_1^2-\sigma_2^2-2 s\right)}{\beta\gamma \sqrt{8+\beta^2}}\,.
\label{eq:defs}
\end{eqnarray}

The solution \eqref{eq:fix_prob} is obtained by integrating \eqref{eq:FP} once, to find the gradient 
\begin{equation}
\label{eq:gradP}
\partial_xP_\text{fix}(x)=\text{const.}\frac{\left(1+\alpha(2x-1)\right)^{\zeta/2-1}}{\left(1-\alpha(2x-1)\right)^{\zeta/2+1}}\,.
\end{equation}
The expression \eqref{eq:gradP} is verified to be proportional to the derivative of $\left(\frac{1+\alpha(2x-1)}{1-\alpha(2x-1)}\right)^{\zeta/2}$ and boundary conditions are then imposed to fix the two constants of integration. The resulting expression is finally transformed into Eq.~\eqref{eq:fix_prob} by using the elementary identity: $2\,\text{Tanh}^{-1}(x)=\log\left[(1+x)/(1-x)\right]$. It is verified that in the limit $\zeta\to 0$, one recovers the expression given in the main text.

All in all, the behavior we discussed in the main text is validated by analyzing the fixation probability: Both a higher growth rate and a smaller variability are beneficial for an individual.

\section{Average time for fixation}
\subsection{Neutral case}
The expression for the time of fixation in the neutral case that we presented in the body of the paper is derived as follows. The average time for fixation obeys the following backward equation,
\begin{eqnarray}
&\left\{ 1\!-\!2x+\left[ x(1\!-\!x)+\!\frac{\gamma}{K\tilde\sigma^2}\right] \partial_x\right\} \partial_x T(x)= \nonumber \\
&-\left(\tilde\sigma^2x\left(1-x\right)\right)^{-1}, \label{eq:FP_time}
\end{eqnarray}
with $\tilde\sigma=(1-\epsilon)\sigma$ the boundary conditions $T(0)=T(1)=0$. Integrating Eq.~\eqref{eq:FP_time} and by variation of constants, we obtain:
\begin{eqnarray}
\partial_xT(x)=\frac{1}{x(1\!-\!x)+\!\frac{\gamma}{K\tilde\sigma^2}}\!\left[A+\frac{1}{\tilde\sigma^2}\ln\left(\frac{1-x}{x}\right)\right]\,,
\label{T_der}
\end{eqnarray}
where $A$ is a constant to be fixed by the boundary conditions. The integrals $\int_0^x$ of Eq.~\eqref{T_der} 
needed for $T(x)$ are performed 
by decomposing the rational function at the prefactor and using the formula\,:
\begin{equation}
\int \frac{\ln\left(a+bx\right)}{x}\,dx=\ln a\ln x-\text{Li}_2\left(-\frac{bx}{a}\right),\,\, a>0\,,
\label{Li}
\end{equation}
that follows from the very definition of the dilogarithm $\text{Li}_2(x)=-\int_0^x\ln\left(1-u\right)/u\,du$ (see~\cite{Zagier:2007}).
The formula \eqref{Li} is used four times either directly (with a simple change of variables) or first integrating by parts to satisfy the 
condition $a>0$ in \eqref{Li}. The resulting expression is then transformed to the form given in the main text (which is the one given by Mathematica) by using the reflection property, $\text{Li}_2(x)+\text{Li}_2(1-x)=\text{Li}_2(1)-\ln x\ln\left(1-x\right)$, see~\cite{Zagier:2007}.   

\subsection{General case}
In the general case when selection is present, the expression for the average fixation time cannot be found explicitly but  is reducible to quadratures as follows. The fixation time obeys the backward equation \eqref{eq:FP} with the left-hand side replaced by $-1$. Using the definitions \eqref{eq:defs}, we obtain
\begin{eqnarray}
&&\left\{ 2\frac{K}{\gamma}(s+\sigma_2^2-\epsilon\sigma_1\sigma_2)-2\beta^2x +\!\left[\beta^2x(1\!-\!x)+2\right] \partial_x\right\}\times\nonumber \\
&& \partial_xT(x)=-\frac{2K}{\gamma x(1-x)}. \label{eq:FPT}
\end{eqnarray}
Boundary conditions are $T(0)=T(1)=0$. The homogeneous solution was already found following \eqref{eq:gradP} and reads 
\begin{equation}
T_{hom}(x)=C_1+C_2\,\left(\frac{\chi_+(x)}{\chi_-(x)}\right)^{\zeta/2}\,,
\label{eq:homo}
\end{equation}
where $C_1$ and $C_2$ are constants and we defined 
\begin{equation}
\chi_+(x)\equiv 1+\alpha(2x-1)\,,\quad \chi_-(x)\equiv 1-\alpha(2x-1)\,,
\end{equation}
to simplify notation. The non-homogeneous solution for the gradient of $T$ is obtained by varying the constant in \eqref{eq:gradP}, remarking that $\beta^2x(1-x)+2=\chi_+(x)\chi_-(x)(8+\beta^2)/4$ and integrating the resulting 
first-order differential equation to obtain
\begin{eqnarray}
\partial_x\,T_{part}(x)=-\frac{K\chi_+(x)^{\zeta/2-1}}{2^{\zeta/2}\gamma\left(\zeta/2+1\right)}\times \nonumber \\
\left[\left(\alpha-1\right)\!F_1\!\left(\frac{\zeta}{2}+1,\frac{\zeta}{2},1;\frac{\zeta}{2}+2;\frac{1}{2}\chi_-(x),\frac{\chi_-(x)}{1+\alpha}\right)+\right.\nonumber \\
\!\!\left.\!\!\left(\alpha+1\right)\!F_1\!\left(\frac{\zeta}{2}+1,\frac{\zeta}{2},1;\frac{\zeta}{2}+2;\frac{1}{2}\chi_-(x),\frac{\chi_-(x)}{1-\alpha}\right)\!\right]\nonumber
\end{eqnarray}
where $F_1$ is the hypergeometric function of two variables \cite{GR}. The solution for $T$ involves the integral $\int_0^x \partial_y\,T_{part}(y)\,dy$ of the expression above (for which a closed form does not seem to be available), and the two constants in \eqref{eq:homo} are fixed by 
\begin{eqnarray}
C_1+C_2\left(\frac{\chi_+(0)}{\chi_-(0)}\right)^{\zeta/2}&=&0\nonumber \\
 C_1+C_2\left(\frac{\chi_+(1)}{\chi_-(1)}\right)^{\zeta/2}&=&-\int_0^1\partial_y\,T_{part}(y)\,dy\,.\nonumber
\end{eqnarray}

It is verified from the expression above or directly from the original equation \eqref{eq:FPT} that in the two 
limits $x\to 0$ and $x\to 1$ the solution behaves like in the neutral case, \emph{i.e.} $-K/\gamma x\log x$ and $-K/\gamma(1-x)\log(1-x)$.  Selection and the rest of the parameters affect of course the solution in the rest of the interval of definition $x\in [0,1]$.

\section{Coexistence time}
\begin{figure}[t]
\includegraphics[width=\columnwidth]{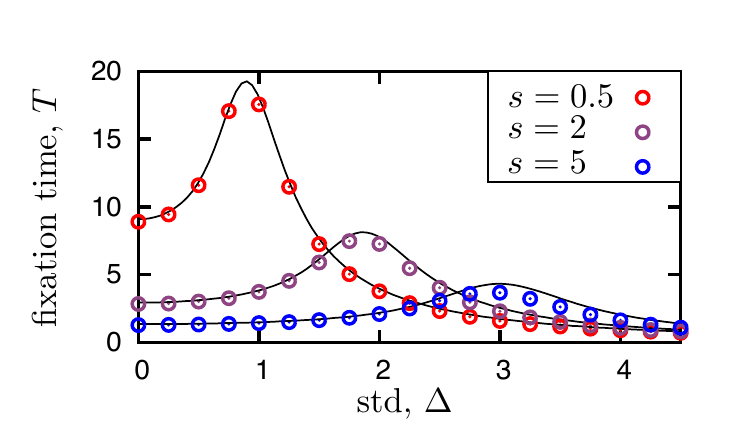}
\caption{\label{fig:coex}Extinction time depending on $\Delta$ for different values of the selection strength: $s=0.5$ (red), $s=2$ (violet) and $s=5$ (blue). Dots are numerical solutions of the Langevin equations,~Eq.~\eqref{eq:Lang}, and black lines are solutions of Eq.~\eqref{eq:FPT}.}
\end{figure}

\begin{figure}[b]
\includegraphics[width=\columnwidth]{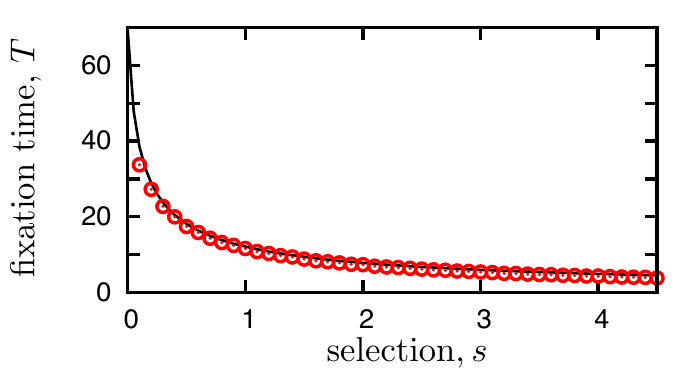}
\caption{\label{fig:FP} Extinction time for different values of $s$ and $\Delta=\sqrt(2s)$. This combination of $s$ and $\Delta$ corresponds to a stable fixed point at $x^*=0.5$ and the maximal coexistence time for each value of $s$, see Fig.~\ref{fig:coex}. As not only the selection strength but also the variability is increasing from left to right, the fixation time is a monotonically decreasing function of $s$.}
\end{figure}

Depending on the position of the stable fixed point, coexistence between two species (one with a larger growth rate, one with a smaller variability, $\nu_1>\nu_2$ and $\sigma_1=\Delta,~\sigma_2=0$) is possible. In this section we present some additional data demonstrating this. In Fig.~\ref{fig:coex}, the extinction time which corresponds to the time of coexistence is shown depending on $\Delta$ is shown for different values of $s$. Dots correspond to solutions of Eqs.~\eqref{eq:Lang} and black lines are numerical solutions of Eq.~\eqref{eq:FPT}. The extinction time has a maximum which exactly coincides with the parameter values of a fixed point $x^*=0.5$. The dependence of this maximal extinction time on the selection strength $s$ is shown in Fig.~\ref{fig:FP}.

 \section{Mapping individual-based models onto the Langevin dynamics}
 The aim of this Section is to show that individual-based models are described by the Langevin equations, Eqs.~\eqref{eq:Lang}, 
 discussed in the main text and to analyze the mapping between the parameters of the two models. 
 
 The environmental conditions change stochastically at the rate $1/\tau$ and are distributed according to a distribution, $p(E)$, with mean  $\langle E\rangle $ and variance $\text{Var[E]}$. The dependency of the instantaneous reproduction rate $\lambda_S(E)$ on $E$ is given by 
 the sigmoidal function\,:
  \begin{align}
 \lambda_S(E)=\phi_S+\omega_S\tanh\left(\frac{\alpha_S E}{2}\right)\,,
 \label{eq:lambda}
 \end{align}
which reduces to $\phi_S\pm\omega_S$ in the limit of large variances $\text{Var[E]}$.  Birth rates are defined as,
\begin{align}
\Gamma^i_\text{repr,S}=\frac{1-m~~~}{1-m^{M^{i}}}\sum_{j=1}^{M^{i}}m^{j-1}\lambda(E^i_j).
\label{eq:Gamma}
\end{align} 
In the no-memory limit $m=0$, the growth rate is therefore given by the instantaneous growth rate $\lambda_S(E)$, while for $m\rightarrow1$ the current growth rate is the arithmetic mean of all previously experienced environments. Death rates are given by $\Gamma^i_\text{death,S}=\gamma N/K$.
\subsection{No memory, $m=0$}
We discuss first the model without memory, where the memory parameter, $m$, is zero\,: Individuals reproduce with the instantaneous reproduction rates [Eq.~\eqref{eq:lambda}], which reduce to $\phi_S\pm \omega_S$ in the limit of large environmental variance. We consider an interval of length $\delta t\gg \tau$ such that the probability for an individual to reproduce or die is small, yet the total number of events occurring over the whole population ($\sim K\gg 1$) is large.
Neglecting the standard demographic noise term \cite{Kimura}, the variation of the $S$-type population is given by
\begin{align}
N_S(t+\delta t)\simeq N_S(t)+N_S(t)\left(\phi_S-\gamma\frac{N(t)}{K}\right)\delta t+\qquad \nonumber\\ 
+\omega_S\int_t^{t+\delta t}N_S(s)\hat{\sigma}(s)\,ds\nonumber\\ \label{eq:stepML}
\end{align}
where $\hat{\sigma}(s)$ is the environmental Boolean random variable that takes values $\pm 1$ and switches with characteristic time $\tau$. The last term of Eq.~\eqref{eq:stepML} is estimated as follows 
\begin{align}
\int_0^{\delta t}N_S(t+s)\hat{\sigma}(t+s)\,ds\simeq N_S(t){\cal G}^e+\nonumber \\
+N_S(s)\omega_S\int_0^{\delta t}\hat{\sigma}(t+s)\,ds\int_0^s\hat{\sigma}(t+s')\,ds'\,,
\label{eq:Strat}
\end{align}
where ${\cal G}^e$ is a Gaussian random variable having zero mean and variance 
\begin{equation}
\text{Var}[{\cal G}^e]=\int_0^{\delta t}ds\int_0^{\delta t}ds'\langle
\hat{\sigma}(s)\hat{\sigma}(s')\rangle=2\tau\delta t\,.
\label{eq:corrsigma}
\end{equation}
Here, we used that $\langle
\hat{\sigma}(t)\hat{\sigma}(t')\rangle=e^{-|t-t'|/\tau}$ and $\delta t \gg \tau$. 
The second term in Eq.~\eqref{eq:Strat} is evaluated 
at the order $\delta t$ using the same integral, Eq.~\eqref{eq:corrsigma}, and gives $N_S(t)\omega_S\tau\delta t$. Combining back all the terms,  we conclude that the equation \eqref{eq:stepML} is equivalent to the Langevin equation~\eqref{eq:Lang} with the mapping of the parameters
\begin{equation}
\nu_S=\phi_S+\omega_S^2\tau\,;\quad \sigma_S^2=2\omega_S^2\tau\,.
\label{eq:mapping}
\end{equation}
Note that the standard demographic noise term in Eq.~\eqref{eq:Lang} should a priori include the fluctuating environmental term $N_S(t)\omega_S{\cal G}^e$ in the sum of the rates. In fact, it can be safely ignored as $\phi_SN_S\delta t\gg \omega_SN_S\sqrt{2\tau\delta t}$ due to 
$\phi_S\ge \omega_S$ and $\delta t\gg \tau$.

\begin{figure}
\centering
\includegraphics[width=0.85\columnwidth]{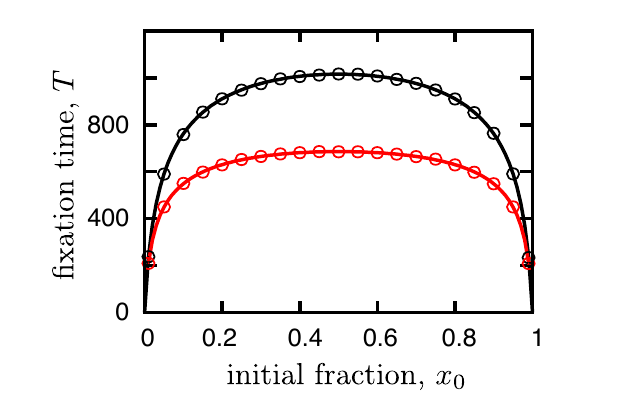}
\caption{Comparison of  data obtained by simulations in the neutral case with fixed (red) and exponentially distributed (black) environmental changes. Both sets of data agree with our analytic calculations, where we used the mappings $\sigma^2=\omega^2\tau$ for fixed times of environmental changes and $\sigma^2=\omega^22\tau$  for exponentially distributed switches. Thereby, the data confirms that the origin of the factor 2 in the mapping is solely the exponential distribution of the environmental changes. Parameters are $\phi_1=\phi_2=1$ and $\omega_1=\omega_2=0.9$, $\gamma=1$, $K=5000$, $\langle E\rangle=0$, $\text{Var}[E]=100$ and $\tau=0.01$.
\label{fig:compare}}
\end{figure}

Finally, the factor $2$ appearing in $\sigma_S^2$ in \eqref{eq:mapping} depends on the Poisson statistics of the environmental fluctuations. If the duration is fixed and equal to $\tau$, Eq.~\eqref{eq:corrsigma} becomes $\tau\delta t$. In that case, the corresponding mappings are $\nu_S=\phi_S+\omega_S^2\tau/2$ and $\sigma_S^2=\omega_S^2\tau$.
This is confirmed numerically in Fig.~\ref{fig:compare} where we show data for exponentially distributed (black) and fixed duration (red) environments. Solid lines are analytic solutions of the fixation time [Eq.~(6) main text] employing the respective mappings.

\subsection{Finite memory, $m>0$}
We now turn to the scenario where memory extends over several environmental conditions that an individual previously experienced. Whilst for $m=0$ an exact analytic mapping can be found in the limit of small $\tau$, for finite memories the situation is more intricate. However, for the special case $m=1$  the variability in the growth rate can be well approximated by the following argument. The reproduction rate $\Gamma^i_\text{repr,S}$ of an individual, $i$, of type $S$ at time $t$ depends on the average of  all instantaneous  reproduction rates $\lambda_S(E)$ experienced by the individual\,:
\begin{align}
\Gamma^i_\text{repr,S}(t)=\frac{1}{M^i(t)}\sum_{k\leq M^i(t)} \lambda^k_S( E_k^{i})=\phi_S+\tilde{\Gamma}^i_S(t)\,.
\label{eq:Gamma}
\end{align} 
At the time of reproduction, we assume for simplicity that offspring looses its memory of past environments experienced by the progenitor. 

\begin{figure}[t]
\includegraphics[width=\columnwidth]{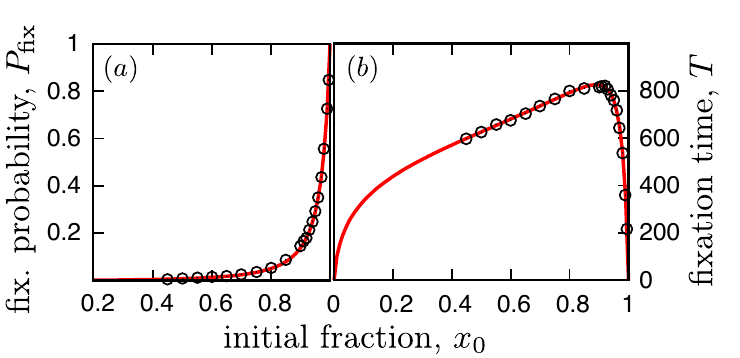}
\caption{Fixation probability and time for non-neutral evolution with memory $m=1$. Black dots correspond to simulation results of the IBM and red lines are analytic solutions. To obtain the latter we fitted  Eq.~\eqref{eq:fix_prob} and the solution of Eq.~\eqref{eq:FPT} to the IBM. We used $\tilde A=s+\sigma_2^2-\sigma_1\sigma_2\epsilon$ and $\tilde B=\sigma_1^2+2\sigma_1\sigma_2\epsilon+\sigma_2^2$ as fitting parameters and obtained $\tilde A=-3.04\times 10^{-3}$ and $\tilde B=8.25\times10^{-3}$. Other parameters are $\phi_1=\phi_2=1$, $\omega_1=0.9$, $\omega_2=0$, $\tau=1/100$, $\gamma=1$, $K=5000$, $\langle E\rangle=0$ and $\text{Var}[E]$.
\label{fig:fit}}
\end{figure}

We consider again a time interval of length $\delta t$ as in \eqref{eq:stepML}. Fluctuations in the rates 
$\tilde{\Gamma}^i_S(t)$ decorrelate on timescales of the order of the lifetimes of individuals, $t_{\text{life}}$, which are much longer than $\tau$ and $\delta t$. Therefore on the $\delta t$ scale, noise is smooth, contrary to \eqref{eq:stepML}.  Conversely, timescales of several lifetimes are much smaller than those on which selection acts and much longer than the characteristic time of the noise. Therefore, to describe the dynamics of the fractions, the environmental noises are well approximated by a shortly correlated noise. An estimation of the 
amplitude of the noise is obtained by calculating the sum 
\begin{equation}
\sigma_S^2 \sim \frac{1}{\tau} \sum_{\ell=-\infty}^{\infty}\sum_{i=1}^{N_S}\sum_{j=1}^{N_S}\langle \tilde\Gamma^i_{S}(t_k)\Delta t_{k} 
\tilde\Gamma^j_{S}(t_{\ell})\Delta t_{\ell}\rangle.\nonumber
\end{equation}
The durations $\Delta t$ of the environmental intervals are independent for different $k$ and $\ell$ (and the contribution $k=\ell$ is negligible with respect to the rest of the sum)
so that one can replace them by $\tau$. In addition, the symmetry in the indices of the intervals allows us to further simplify the expression
\begin{eqnarray}
\sigma_S^2&\sim&2\tau \sum_{\ell=k}^{\infty}\sum_{i=1}^{N_S}\sum_{j=1}^{N_S}\langle \tilde\Gamma^i_{S}(t_{k}) \tilde\Gamma^j_{S}(t_\ell)\rangle\,.
\label{eq:OmegaII}
\end{eqnarray}
To compute the average in Eq.~\eqref{eq:OmegaII} three different orders of events have to be distinguished: a) 
If the birth of the $j$-th  individual was prior to the one of the $i$-th 
individual, then it follows from Eq~\eqref{eq:Gamma} that the quantity to be averaged is $\omega_S^2M^i_{k}/\left(M^i_{k}M^j_{\ell}\right)=\omega_S^2/M^j_{\ell}$, where $M^j_{\ell}$ is the number of environmental switches since the birth of the $i$-th individual up to time $t_k$. To derive Eq.~\eqref{eq:OmegaII} we have used that the terms in the sum~\eqref{eq:Gamma} take independent values $\pm \omega_S$ with equal probability. Conversely, case (b) is when  the birth of the $j$ individual is posterior to the one of the $i$-th 
individual. Then the quantity to be averaged is $\omega_S^2\left(M^i_{k}-\delta b\right)/\left(M^i_{k}M^j_{\ell}\right)$, where $\delta b$ is the time between the birth of the $i$-th and the $j$-th individuals. Finally, in case (c) when $\delta b>M^i_{k}$,  the 
correlation is zero as there is no overlap between the environmental fluctuations of the two individuals.
Due to the Poissonian nature of the events, the number of switches since birth (back in the past) or before reproduction (forward in the future) have the same distribution $\exp(-M/\overline{M})/\overline{M}$ where $\overline{M} \sim t_{\text{life}}/\tau$ (its exact value does not affect the sequel). It follows that 
\begin{eqnarray}
\sigma_S^2 \sim \frac{2\omega_S^2\tau}{\overline{M}^2}\left[
\int_0^\infty \!\!\!\!dt\int_0^{\infty}\!\!\!\!du\int_0^\infty \!\!\!dv\, e^{-\left(t+2u+v\right)/\overline{M}}\frac{1}{u+v+t}\right.\nonumber \\
\left. +\int_0^\infty dt\int_0^{\infty}du\int_0^u dv\, e^{-\left(t+2u-v\right)/\overline{M}}\frac{u-v}{\left(u-v+t\right)u}\right]\,.\nonumber\\
\label{eq:mess}
 \end{eqnarray}
The integral over $t$ is the continuous approximation of the sum over $\ell-k$ appearing in Eq.~\eqref{eq:OmegaII} while 
the variables $u$ and $v$ refer to the variables $M^i_{k}$ and $\delta b$.
 The first and second term in the square parentheses of \eqref{eq:mess} correspond to cases (a) and (b), respectively. 
By a series of change of variables and integrations by parts, it is shown that \eqref{eq:mess} reduces to 
\begin{equation}
\sigma_S^2 \sim \frac{2\omega_S^2\tau}{\overline{M}^2}\int_0^{\infty}du\,e^{-u/\overline{M}}\int_0^u dv\,e^{-v/\overline{M}}=\omega_S^2\tau\,.
\label{eq:finmap}
\end{equation}
The validity of this approximation is confirmed for the neutral case in Fig.~2 in the main body of the paper where the fixation probability and time are compared to the analytic calculations employing Eq.~\eqref{eq:finmap}. Additional data for non-neutral evolution is presented in Fig.~\ref{fig:fit} where two species (one with finite variability, $\omega=0.9$, one with vanishing variability) are analyzed. Analytic solutions for the fixation time and probability are fitted to simulation data. The best fit deviates less than $1\%$ from Eq.~\eqref{eq:finmap}. 



\begin{figure}[b]
\includegraphics[width=0.9\columnwidth]{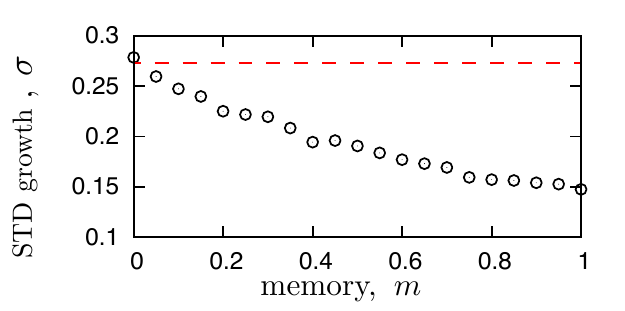}
\caption{Numerical estimation of the STD of the noise dependence on the memory parameter $m$. Extinction times in the neutral scenario were measured. By evaluating the inverse of the fixation time function [Eq.~(6) main text] the STD of the growth was calculated. For $m=0$ the result agrees with the mapping, cf. Eq~\eqref{eq:mapping}, indicated by the red dashed line. For larger values of $m$ the variability is reduced. Parameters are $\phi_1=\phi_2=10$, $\omega_1=\omega_2=8.237$, $\tau=1/500$, $\gamma=1$, $K=100$, $\langle E\rangle=0$ and $\text{Var}[E]$. \label{fig:sigma}}
\end{figure}

We now briefly discuss the dependence of $\sigma_S$ on the memory parameter $m$. In Fig.~\ref{fig:sigma}, the STD of the noise in the growth rate, $\sigma$ ,depending on the memory parameter, $m$, is analyzed. Results were obtained by simulating the neutral evolution case, measuring the fixation time and calculating $\sigma$ employing the analytic expression for the extinction time [Eq.~ 6 main text]. For $m=0$ the thereby obtained value agrees nicely with the mapping introduced above indicated by the red dashed line [Eq.~\eqref{eq:mapping}]. With increasing memory, $m$,  the STD of the noise, $\sigma$, decreases monotonically and approaches Eq.~\eqref{eq:finmap} for $m\rightarrow1$.

\begin{figure}[t]
\includegraphics[width=0.9\columnwidth]{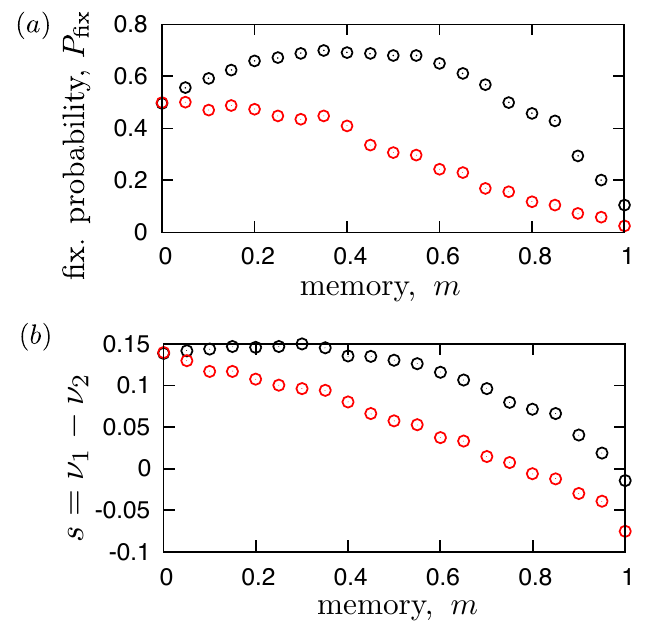}
\caption{(a) Fixation probability  and (b) selection coefficient depending on the memory parameter $m$. The first species' growth rate depends on the environment  while the second one's is constant. Black dots correspond to the IBM introduced in the main body of the paper, while red dots represent a model modification not memorizing the first environment (the one in which an individual is born; for details see text). While the fixation probability is a direct simulation result, the selection coefficient $s$ is inferred from it using the additional data presented in Fig.~\ref{fig:sigma}. Parameters are  $\phi_1=\phi_2=10$, $\omega_1=8.237$, $\omega_2=0$, $\tau=1/500$, $\gamma=1$, $K=100$, $\langle E\rangle=0$ and $\text{Var}[E]$. \label{fig:kernel_sel}}
\end{figure}

Let us now analyze the mapping of the average reproduction rate $\nu_S$. Importantly, this mapping is very sensitive to model details which we will exemplify in the following.  As results for neutral species do not dependent on the average reproduction rate, we have to turn to  the evolution of non-equal individuals to understand the mapping of the growth rates. In Fig.~\ref{fig:fit}, we show the fixation probability for two species with the same $\phi_1=\phi_2=1$ but only the first species has a variable reproduction rate $\omega_1=0.9,~\omega_2=0$ for $m=1$. Red lines correspond to a fit with $s=\nu_1-\nu_2\approx-0.0030$ and agree perfectly with simulation results. In other words, the first species does not only have a disadvantage due its sensitivity on environmental changes, $\sigma_1>\sigma_2$, but  also has a smaller average growth rate. To study this effect in more detail, let us now  analyze the fixation probability dependence on the memory parameter $m$, see Fig.~\ref{fig:kernel_sel} panel (a). Black dots correspond to the standard IBM (if not mentioned otherwise our discussion applies to this data), red dots to a slightly changed model which is going to be introduced in the following. For $m=0$ both species are equally likely to fixate as the growth advantage of the more variable species $\nu_1=\phi+\omega_1^2\tau>\nu_2=\phi$ exactly compensates for its disadvantage due to the STD of the noise $\sigma_1>\sigma_2$.
 For increasing values of $m$ first the more variable ($m<0.7$) later the less variable species is favored ($m>0.7$). Whether this behavior is caused by the STD of the noise or differences in the mean reproduction rates is not obvious as the influence of both fitness contributions is of  comparable strength. Therefore, we estimate the selection coefficient, $s=\nu_1-\nu_2$, from the fixation probability data, see  Fig.~\ref{fig:kernel_sel} panel (b). This is achieved by assuming that the variability of species {\bf 1} with $\omega_1=0$ is zero ($\sigma_1=0$) and that the variability of species {\bf 2} with $\omega_2=8.237$ is the  same as in the neutral evolution scenario and thereby given by the data presented in Fig~\ref{fig:sigma}. Note that this approximation might neglect some higher noise correlations arising due to the coupling of both species via the carrying capacity. For $m=0$ the thereby obtained value of $s$ agrees well with our analytic results, cf. Eq.~\eqref{eq:mapping}. With increasing $m$ the growth rate of the more variable species is decreased till the selection coefficients becomes negative effectively favoring the more variable species. However the decrease of the selection coefficient with $m$ is smaller than the reduction of $\sigma$ shown in Fig.~\ref{fig:sigma}. Hence, for small $m$ the more variable species is favored as its advantage due to a larger average reproduction rate is larger than its disadvantage due to its sensitivity on the environment. This advantage in the growth rate is more sensitive to details in the IBM in comparison to the variability discussed in the main text for the Langevin equation Eq.~(1).
We are going to illustrate this by analyzing a slightly modified version of the IBM. But before doing so, we present an intuitive argument explaining one factor influencing the average growth rate: When an individual is born it experiences the current environment shorter than the average length of an environment. However, the model weights all experienced environments equally, see Eq.~(7) in the main text. As the first experienced environment is more likely to be a good environment [more reproduction events happen during more beneficial environments], higher growth rates have a larger weight in the average and the average growth rate of the variable species is effectively increased. To obtain a description including this factor, it would be best to perform a time average over all previously experienced environments. Unfortunately, such a procedure is computationally very expensive. We therefore, test our explanation for the bias by not including the very first environment, the one in which an individual is born, in the averaged reproduction rate. The red dots in Fig.~\ref{fig:kernel_sel} corresponds to simulation results for this modified model. Even though all parameters are the same and for $\tau=1/500$ and $\phi\approx10$ an individual experiences in average 25 different environments, the small modification of the model substantially changes the simulation results. While the modification almost has no impact on the STD of the noise, it alters the average reproduction rate. For instance the regime in which the more variable species is favored completely disappears, cf. red dots Fig.~\ref{fig:kernel_sel} where $P_\text{fix}\leq0.5$ for all $m$. This example illustrates that on the first sight tiny details of an IBM might substantially influence the evolutionary outcome and that one should be cautious when drawing conclusions from them. Importantly, the mechanism discussed in the main text does not rely on specific assumptions of the microscopic models: A finite STD of the growth rate is always a disadvantage. It might be compensated for by a larger average reproduction rate but the same value of the growth rate without variability is always preferable.

\section{DEPENDENCE ON THE SWITCHING RATE}

 \begin{figure}[t,b]
 \includegraphics[width=\columnwidth]{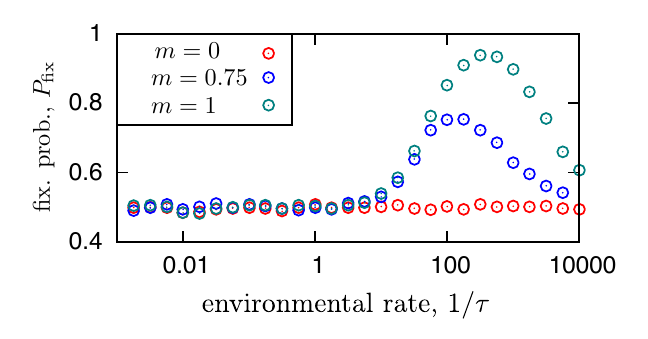}
 \caption{\label{fig:compare_models} 
Dependence of the fixation probability on the environmental switching rate. Both species have the same $\phi_1=\phi_2=10$, but only the second species is sensitive to the environment ($\omega_1=0,~\omega_2=9$).  For no memory ($m=0$), both species are equally likely to fixate, as the advantage in the average growth rate of species {\bf 2} exactly compensates for its disadvantage due to its sensitivity on the environment. For $m>0$ those two effects do not cancel out anymore and the second species is favored. Other parameters are $\gamma=1$, $K=100$, $\alpha=1$, $\langle E\rangle=0$ and $\text{Var}[E]=100$.}
 \end{figure}
 
In this Section, we present additional data for the non-neutral case.  Fig.~\ref{fig:compare_models} shows the fixation probability depending on the environmental switching rate $1/\tau$. In particular, we investigate extinction for a species which is not sensitive to its environment ($\phi_1=\phi=10$, $\omega_1=0$) competing with a sensitive species ($\phi_2=\phi=10$, $\omega_2=9$) for different values of $m$. In the case of no memory $m=0$ both species are equally likely to fixate as the advantage in the average reproduction rate $\nu_2=\phi+\omega_2\tau$ exactly compensates for the disadvantage due to the STD of the noise in the growth rate [see Eq.~\eqref{eq:mapping}]. For larger values of the memory parameter, a bias favoring the species with $\omega=0$ is present (the exact value of the fixation probability depends on mapping details as discussed above).  Importantly, the bias is not only present for very quickly fluctuating environments, but already emerges if reproduction events happen on a time scale comparable to $\tau$. This supports the conclusion, that we were already drawing in the body of this paper when discussing Fig.~4: the white noise approximation is an adequate description for such an evolutionary process in that parameter regime.


\begin{thebibliography}{28}
\bibitem[]{FN}\bibinfo{misc}{For simplicity we restrict our discussion to the situation where the environment only affects the birth rate; the influence on the death rate is analyzed  analogously and leads to the same conclusions.}
\bibitem[]{FN2}\bibinfo{misc}{Without loss of generality, we assume that larger values of $E$ correspond to better environmental conditions.}
\end{thebibliography}
\end{document}